\documentclass{article}
\usepackage[sort]{natbib}
\usepackage{graphicx} 
\usepackage[dvips]{epsfig}
\def\d{{\rm d}}
\def\i{{\rm i}}

\def\eta{{r}}

\title{Stability of flow through a slowly 
diverging pipe}

\author{Kirti Chandra Sahu, Rama Govindarajan \\ 
Engineering Mechanics Unit, Jawaharlal Nehru Centre \\ for 
Advanced Scientific Research, Bangalore 560 064, India. \\
E-mail: rama@jncasr.ac.in}
\begin{document}
\maketitle
\begin{abstract}
Although the critical Reynolds number for linear instability of the laminar 
flow in a straight pipe is infinite, we show that it is finite for a divergent
pipe, and approaches infinity as the inverse of the divergence angle. The
velocity profile at the threshold of inviscid stability is obtained. 
A non-parallel analysis yields linear instability at surprisingly low 
Reynolds numbers, of about $150$ for a divergence of $3$ degrees, which would 
suggest a role for such instabilities in the transition to turbulence. A 
multigrid Poisson equation solver is employed for the basic flow, and an 
extended eigenvalue method for the partial differential equations describing 
the stability.
\end{abstract}

\section{Introduction}
The laminar flow through a straight pipe is linearly stable for any Reynolds 
number [\cite{daveydrazin}, \cite{martin}], but, as first
demonstrated by \cite*{reyn}, 
transition to turbulence usually occurs at a Reynolds number $Re$, based on the
pipe diameter and mean velocity, of around $2000$. By reducing the external 
disturbances, it is possible to achieve laminar flow up to 
$Re \sim 10^5$ [\cite{mann}] when the pipe is smooth and the flow at 
the inlet very quiet. The Reynolds number up to which it is possible to keep
the flow laminar varies inversely as the level of external disturbance
[\cite{mullin1}]. Although questions remain about the complete 
route to turbulence in a straight pipe, it seems likely 
that the spectrum of linear (stable) modes has a role to play 
via transient algebraic growth [\cite{trefethen, schmid1}] of 
disturbances. It has recently been demonstrated both theoretically and 
experimentally that a nonlinear self-sustaining mechanism leads to 
the existence of travelling waves (and time-periodic states) that 
appear to play a key role in shear turbulence 
[\cite{waleffe1, waleffe3, eckh03, eckh04, hof04}].

Our purpose in this paper is to examine the possible role of small local
divergences in the transition process. These could have a large effect since
linear stability is described by a singular perturbation problem.
Whearas a large amount of work has been done on the 
flow in significantly converging/diverging [e.g. \cite{floryan}] and in 
suddenly expanding geometries, we know of no work on the
stability of {\em slowly} diverging pipe flows. Sudden expansions have attracted
attention because of the recirculation zone they generate. In particular,
\cite{fearn} and \cite{cherdron} study the length of the recirculation
zone as a function of the Reynolds number, and \cite{sreenivasan} examine
the oscillations of the recirculating bubble and their effect on the flow.
Our focus is different, as will become clear below. 
\cite*{eagle} and {\cite*{eagle3}}, analysed the 
Jeffery-Hamel flow generated by a slowly diverging channel, and showed by linear
parallel stability analysis (the Orr-Sommerfeld equation), that the critical 
Reynolds number decreases by a large amount even for a small divergence angle.
The divergent {\em pipe} is more interesting for several reasons: the critical
Reynolds number is infinite for an angle of divergence of zero, the Reynolds
number is a decreasing function of the streamwise (axial) coordinate $x$, 
and the flow non-parallelism is larger for a given divergence.
In the present work, we employ a two-pronged approach. For the mean flow, we
derive an axisymmetric Jeffery-Hamel equation (AJH), which is valid at small 
divergence angles.
At larger angles of divergence ($1^\circ$ or greater) we solve the 
Navier-Stokes equations directly in the axisymmetric geometry shown in figure 
\ref{geometry}, with a divergent portion of finite extent. At small angles of 
divergence and high Reynolds numbers (above $1000$), a parallel flow stability
analysis is conducted on the AJH profile, while at lower Reynolds 
numbers, the partial differential equations for non-parallel stability are 
solved as an extended eigenvalue problem.

Our main results may be summarised as follows. 
At low levels of divergence, linear stability is determined by a 
parameter $S(x)$, defined as the product of $Re$ and the slope $a$ 
of the wall. The basic AJH flow is completely described by this parameter.
The flow is unstable to the swirl mode for $S>10$, so 
the critical Reynolds number approaches infinity as $1/a$.
At divergences greater than $1^\circ$, non-parallel effects 
are found to be quite large, and a non-parallel analysis shows that the flow in 
a geometry containing a $3^\circ$ divergence is linearly unstable to the swirl 
mode at Reynolds numbers as low as $150$. 
The following two sections describe the basic flow computations and the
stability analysis respectively.

\section{The basic flow}
\subsection{Axisymmetric Jeffery-Hamel equation}
We begin by noting that unlike in a divergent two-dimensional
channel, no
similarity flow is possible in a divergent pipe. At very low
angles of
divergence, however, it is possible to derive a one-parameter
family of
velocity profiles, where the parameter
\begin{equation}
S \equiv a Re
\label{stabpar}
\end{equation}
varies slowly in the axial coordinate $x$. Here the slope
\begin{equation}
a \equiv {\d R(x_d) \over \d x_d} \ll 1,
\end{equation}
$R$ is the radius of the pipe, and the subscript $d$ stands for a
dimensional
quantity. 
Upon eliminating the pressure from the axisymmetric momentum 
equations, we obtain
\begin{equation}
-{V_d \over r_d}{\partial U_d \over \partial r_d}+U_d {\partial^2 U_d \over 
\partial x_d \partial r_d}+V_d {\partial^2 U_d \over \partial r_d^2}-\nu \left [ 
{\partial^3 U_d \over \partial r_d^3}-{1 \over r_d^2} {\partial U_d \over 
\partial r_d}+{1 \over r_d} {\partial^2 U_d \over \partial r_d^2} \right ]=0,
\label{seq1}
\end{equation}
where $\nu$ is the kinematic viscosity, and $r$ is the radial coordinate.
The axial and radial velocities, $U_d$ and $V_d$ respectively, may be 
written in terms of a generalised streamfunction $\Psi_d$ as
\begin{equation}
U_d={1 \over r_d}{\partial \Psi_d \over \partial r_d}, \qquad
V_d=-{1 \over r_d}{\partial \Psi_d \over \partial x_d},
\end{equation}
to satisfy continuity. Substituting this in equation (\ref{seq1}), we get
$$
3 r_d {\partial \Psi_d \over \partial x_d} {\partial^2 \Psi_d \over 
\partial r_d^2} -3{\partial \Psi_d \over \partial r_d}{\partial \Psi_d 
\over \partial x_d}+{r_d^2}{\partial \Psi_d \over \partial r_d}
{\partial^3 \Psi_d \over \partial r_d^2 \partial x_d}- r_d {\partial \Psi_d 
\over \partial r_d}{\partial^2 \Psi_d \over \partial r_d \partial x_d}
-r_d^2 {\partial \Psi_d \over \partial x_d}{\partial^3 \Psi_d \over 
\partial r_d^3} -
$$
\begin{equation}
\nu \left [r_d^3 {\partial^4 \Psi_d \over \partial r_d^4} -
2 r_d^2 {\partial^3 \Psi_d \over \partial r_d^3}+3 r_d 
{\partial^2 \Psi_d \over \partial r_d^2}-3 {\partial \Psi_d \over 
\partial r_d}\right]=0.
\end{equation}
The above equation is non-dimensionalised using the local radius $R(x_d)$ and 
the centreline velocity $U_c(x_d)$ as scales, e.g., $\Psi_d = U_cR^2\Psi$. 
In particular,
\begin{equation}
\d x_d=R \d x.
\label{nond}
\end{equation}
The Reynolds number is assumed to be high and the divergence small 
enough, so that all terms of $O(Re^{-2})$ or $O(a^2)$ and higher 
are negligible. The resulting equation in non-dimensional form 
becomes
$$
\left[q \Psi- {\eta a \Psi^{\prime}} \right] \left[-3 \Psi^{\prime}  + 3 
\eta \Psi^{\prime \prime}  -{\eta^2} \Psi^{\prime\prime\prime}\right]-\eta 
\Psi^{\prime} \left[q \Psi^{\prime} - \eta \Psi^{\prime\prime} a  - 
\Psi^{\prime} a \right] + \eta^2 \Psi^{\prime} 
$$
\begin{equation}
\left[ ( q - 2 a ) \Psi^{\prime\prime} - \eta a 
\Psi^{\prime\prime\prime}\right]-{1 \over Re} \left[\eta^3 \Psi^{iv} -
2 \eta^2 \Psi^{\prime\prime\prime} + 3 \eta \Psi^{\prime\prime} - 3 
\Psi^{\prime}\right]=0,
\label{similarity2}
\end{equation}
where 
$$q \equiv {1 \over U_cR^2}{d \over d x_d}(U_cR^2),$$
the Reynolds number is defined as $Re(x_d) \equiv U_c(x_d) R(x_d) /\nu$, 
and the primes refer to differentiation with respect to $r$.
For the case of near-similar flow, given constant mass flow rate, we may set
$q=0$, and the above equation may be integrated once with respect to
$r$ to give
\begin{equation}
{\eta^2}U^{\prime\prime\prime} = -\eta U^{\prime\prime} + (1 - 4 S \eta^2 U) 
U^{\prime},
\label{similarity}
\end{equation}
where $U=\Psi^{\prime}/\eta$. The boundary conditions are $U=0$ at $r=1$, 
and $U=1,\hspace {1mm} U'=0$ at $r=0$. Profiles obtained from equation 
(\ref{similarity}) are compared to those obtained from a numerical
axisymmetric Navier-Stokes solution in the following subsection.
\begin{figure}
\begin{center}
\includegraphics[width=0.45\textwidth]{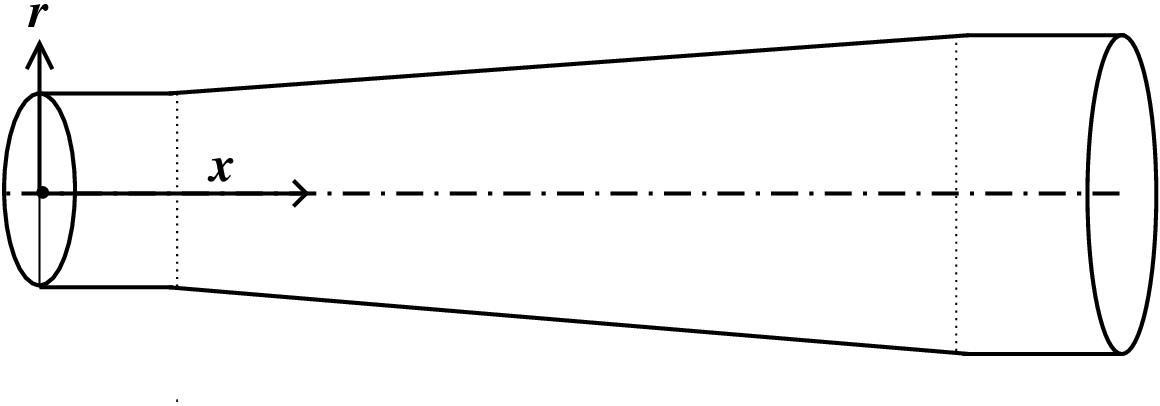}
\caption{Schematic diagram of the divergent pipe used in the numerical 
simulations, not to scale.}
\label{geometry}
\end{center}
\end{figure}

\subsection{Numerical solution}
The geometry studied here is as shown in the schematic in figure 
\ref{geometry}, with a straight pipe at the entry, followed by an axisymmetric
divergent portion, which in turn is followed by a long straight exit 
portion. The length and velocity scales, redefined for convenience in this 
subsection alone, respectively are the radius $R_i$ and the centerline 
velocity $U_i$ at the inlet.
In the case presented here, the divergent portion starts at $x=9.4$ 
and ends at $x=91$ with a $3^\circ$ angle of divergence. The total length of the 
domain is $L=120$.
The axisymmetric Navier-Stokes equations for steady, incompressible, 
Newtonian flow in the streamfunction vorticity formulation, in non-dimensional
form, are given by
\begin{equation}
{\frac {\partial \Omega } {\partial t}} +({\vec{U} . \nabla})
\Omega={1 \over Re_i} {\nabla^2} \Omega,
\label{omegaeq}
\end{equation}
\begin{equation}
\Omega=-{\nabla^2} {\Psi},
\label{psieq}
\end{equation}
where $Re_i \equiv U_i R_i/\nu$, $\Omega(x,r)$ is the azimuthal vorticity,
$\vec{U}$ is the velocity vector, and $t$ is time.
The boundary conditions at the centerline are $\Psi=\Omega=V=\partial
U/\partial r=0$. No-slip and impermeable boundary conditions are 
imposed at the wall. The functional forms of the streamfunction at the 
centerline, and the vorticity at the wall, are described by employing
fictitious points outside the domain. At the inlet, a parabolic velocity 
profile is prescribed, while at the outlet the Neumann boundary conditions:
$\partial \Psi / \partial x = 0$, and $\partial \Omega / \partial x = 0$
are imposed. The reason for using a long exit section, and the consequent
increase in computational effort, is due to the requirement that the
Neumann condition be valid at the exit. 

We begin with a guess solution, where the velocity profile is
parabolic at every axial location, and march in psuedo-time until 
a steady-state solution is obtained. The vorticity distribution at each new 
time step is calculated from (\ref{omegaeq}), adopting a first-order accurate 
forward 
differencing in time and second-order accurate central differencing in space,
on a $34 \times 514$ grid. The vorticity distribution thus computed is used to 
solve the Poisson equation (\ref{psieq}) by a Jacobi iterative scheme to obtain 
the streamfunction everywhere. Numerical acceleration is achieved by a 
six level full-multigrid technique [\cite{cfd1}].
The procedure is repeated until the cumulative change in vorticity during the
time step reduces to below $=10^{-10}$.

\begin{figure}
\centering
\begin{minipage}{1.0\textwidth}
\includegraphics[width=0.45\textwidth]{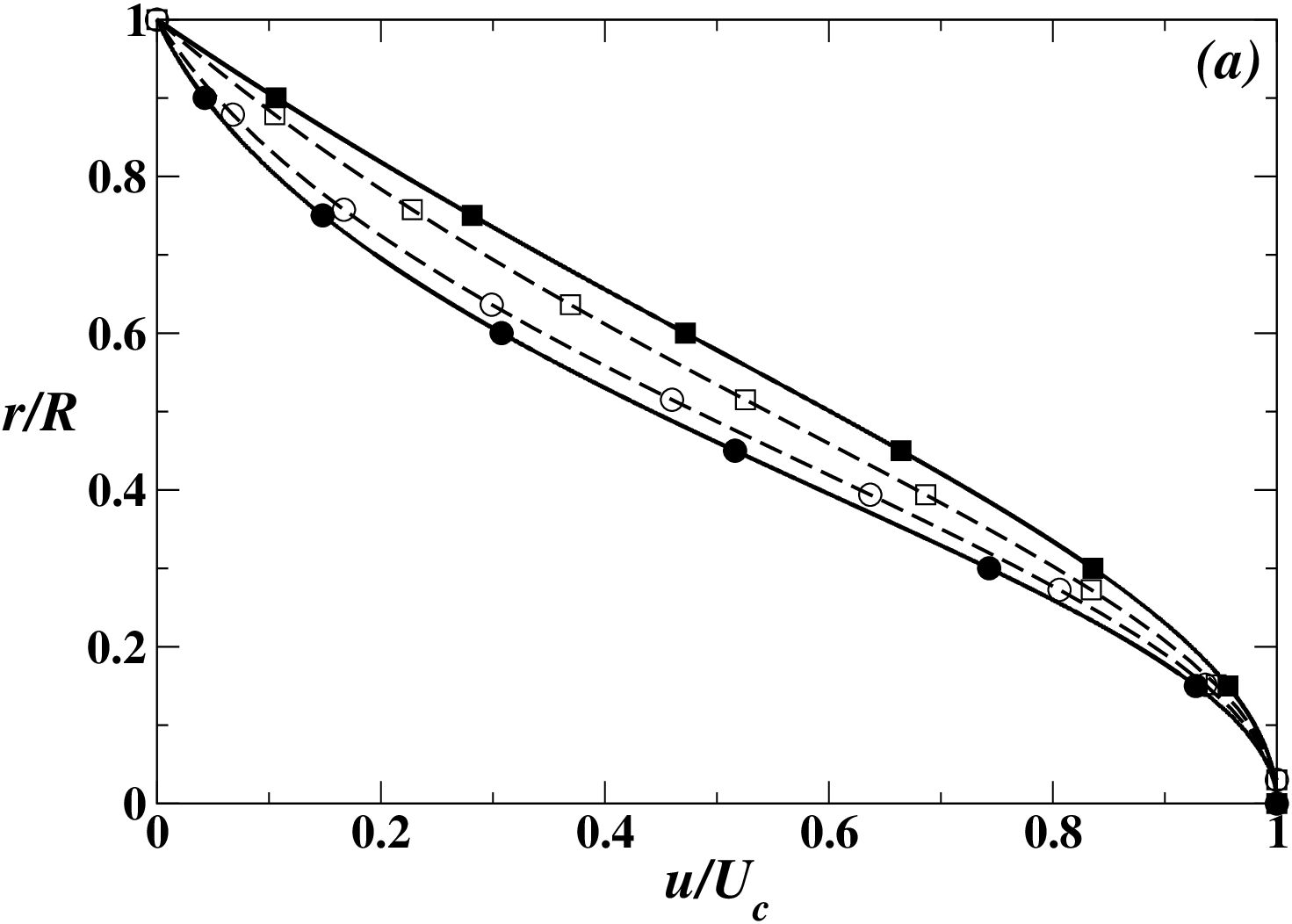}
\hspace{5mm}
\includegraphics[width=0.45\textwidth]{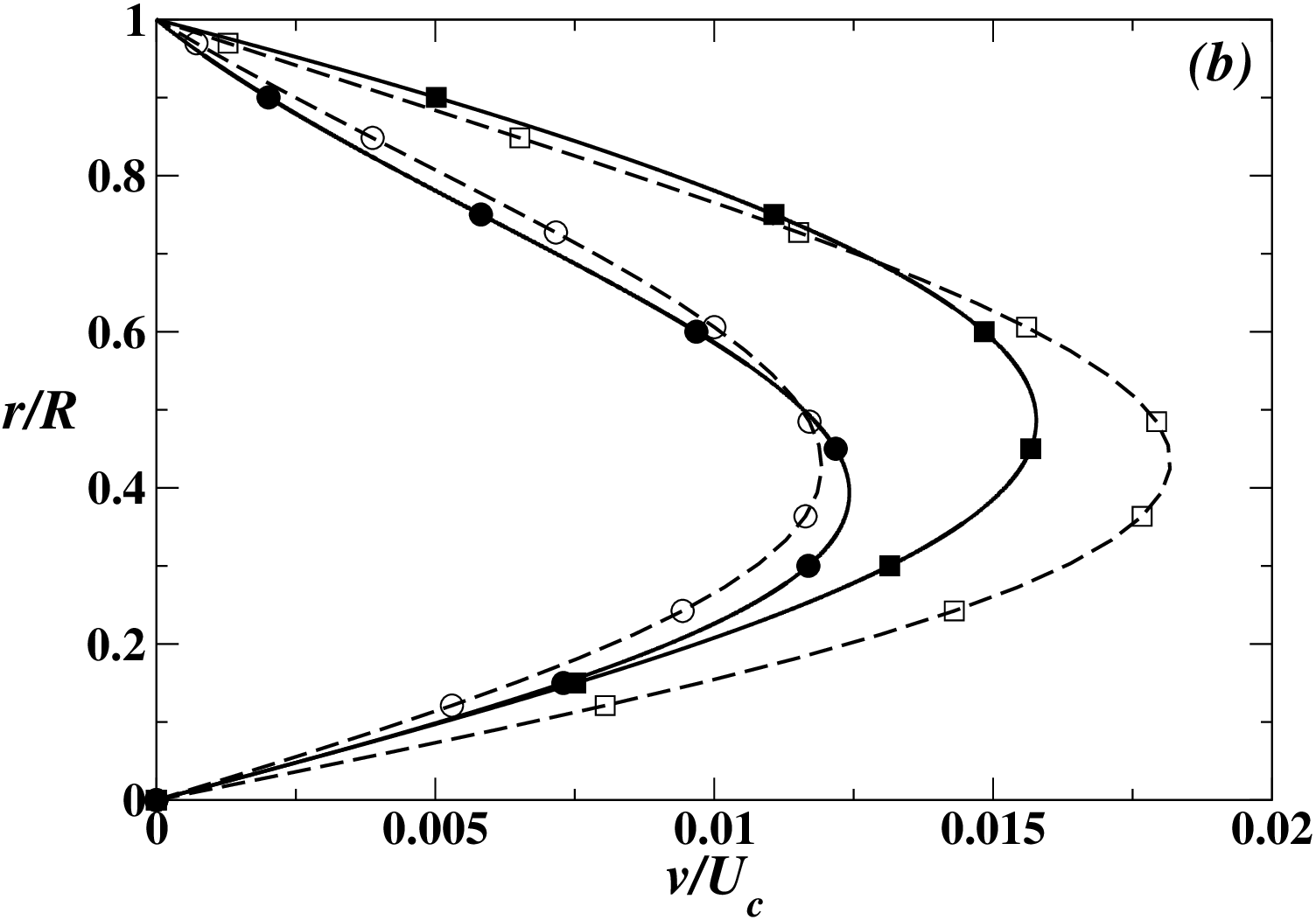}
\caption{Comparison of numerically obtained velocity profiles (solid lines) 
at different axial locations (circles: $x=22.9, S=7.49$ and squares: $x=46.4,
S=3.75$) with the AJH profiles (dashed lines). (a) Axial velocity. 
(b) Radial velocity.}
\label{vel}
\end{minipage}
\end{figure}

The axial and radial velocity profiles for $Re_i=150$ and angle of 
divergence $3^\circ$ at different 
downstream locations are compared with those from the AJH profiles 
in figure \ref{vel}(a) and \ref{vel}(b) 
respectively. It is seen that the AJH profile underpredicts the effect 
of divergence at $x=22.9$ but overpredicts it at $x=46.4$. It is relevant to
mention that the AJH profiles do not correspond to flow through any particular 
geometry. At very low angles of divergence, axial variations are slow, the flow 
attains near-similarity within a short downstream distance, and the AJH profiles
are expected to predict the real flow well. We find this to be the case at
angles of divergence less than a degree.
The profiles computed here are used in the stability 
calculations, as described in the next section. Incidentally, at higher
divergence, regions of separation are obtained to very 
good accuracy by the present method, but are not the subject of discussion
here. To the contrary, our interest is in finding the smallest divergence at
which flow behaviour is completely different from that in a straight pipe.

\section{Non-parallel stability analysis}
We now revert to the use of the local radius $R(x)$ and the local centerline 
velocity $U_c(x)$ at a given $x$ as scales. Each flow quantity is expressed 
as the sum of a steady mean and a time-dependent perturbation, such as
\begin{equation}
u = U(x,r) + \hat u(x,r,\theta,t).
\label{split}
\end{equation} 
Since the flow under consideration varies significantly in the axial
direction, a normal mode form may be used only in time and in the azimuthal
coordinate $\theta$. In the axial coordinate, the perturbation may be
expressed as a rapidly varying wave-like part scaled by a relatively slowly
varying function [see e.g. \cite{Bertolotti,rn}], such as
\begin{equation}
[\hat u, \hat v, \hat w, \hat p] = {\rm Real} \left\{[u(x, r), v(x, r), 
w(x, r), p(x, r)] \exp\left[\i \left(\int{\alpha(x) \d x} + n \theta - 
\beta_d t_d\right) \right]\right\},
\label{normal_mode}
\end{equation} 
where $\hat u$, $\hat v$ and $\hat w$ are the axial, radial and the 
azimuthal velocity perturbations respectively, $\hat p$ is the 
pressure perturbation, $\alpha(x)$ is a local axial 
wavenumber, $n$ is the number of waves in the azimuthal direction, and $\beta$
is the disturbance frequency. Flow quantities in the form (\ref{split}) are
substituted in the Navier-Stokes and continuity equations, the mean flow equation
is subtracted, and nonlinear terms in the perturbations are neglected. 
Since axial variations are slow and the Reynolds number is large, a
consistent approximation is to retain all terms up to $O(a)$ and 
$O(Re^{-1})$ (in the critical and wall layers and elsewhere in the pipe) and 
neglect higher order effects. 
The result is a set of partial differential equations for the perturbation 
velocities and pressure, each of first order in $x$ and up to second order in 
$r$, which amounts to a seventh order system in $r$. These may be expressed 
in the form
\begin{equation}
{\cal H} \phi(x,r) + {\cal G} {\partial \phi(x,r) \over \partial x} = 
\beta {\cal B} \phi(x,r).
\label{npe}
\end{equation}
Here $\phi = [u, v, w, p]$, $\beta=\beta_d R/U_c$ and the non-zero elements of 
the $4 \times 4$ matrix operators ${\cal H}$, ${\cal G}$ and ${\cal B}$ are 
given by
$$h_{11}= U \Big [ 2 {U_c^\prime \over U_c} + \i \alpha  - 
a r {\partial \over \partial r} \Big ]+ {\partial U \over \partial x}-
ar {\partial U \over \partial r}+ V {\partial \over \partial r}+ 
{1 \over Re} \Big [ \alpha^2 +{n^2 \over r^2} -{1 \over r}{\partial 
\over \partial r} - {\partial^2 \over \partial r^2} \Big ],
$$
$$
h_{12} = {\partial U \over \partial r}, \hskip2mm
h_{14} = \Big (2 {U_c^\prime \over U_c} + \i \alpha -ar 
{\partial \over \partial r}\Big ),$$
$$
h_{22} = V {\partial \over \partial r}+ {\partial V \over \partial r}+
U \Big [ {U_c^\prime \over U_c}+ \i \alpha  - ar{\partial \over
\partial r} \Big ] -{1 \over Re} \Big [{\partial^2 \over \partial r^2}+
{1 \over r} {\partial \over \partial r}-{(1+n^2) \over r^2} - \alpha^2 \Big ],
$$
$$
h_{23} = {2 \over Re} { \i n \over r^2}, \hskip2mm 
h_{24} = {\partial \over \partial r}, \hskip2mm 
h_{32} = -{2 \over Re} { \i n \over r^2},$$
$$
h_{33} = V {\partial \over \partial r}-{V \over r}+U \Big [{U_c^\prime 
\over U_c}+ \i \alpha - ar{\partial  \over \partial r} \Big ]
-{1 \over Re} \Big [{\partial^2 \over \partial r^2}+
{1 \over r} {\partial \over \partial r}-{(1+n^2) \over r^2} - \alpha^2 \Big ],
\hskip2mm 
h_{34} = {in \over r},
$$
$$
h_{41} = {\i \alpha \over Re} {\partial \over \partial r},\hskip2mm 
h_{42} = V {\partial \over \partial r} +{\partial V \over \partial r} +
U \Big ({U_c^\prime \over U_c} + \i \alpha - a r {\partial \over \partial r} 
\Big )+
{1 \over Re} \Big ({n^2 \over r^2}+\alpha^2 \Big ),
$$
$$ 
h_{43} = {\i n \over Re} \Big ({1 \over r^2}+{1 \over r} {\partial \over \partial r}\Big ),
\hskip2mm
h_{44} = {\partial \over \partial r}; \qquad
g_{11} = g_{22} = g_{33} = g_{42} = U, \hskip2mm g_{14} = 1, 
$$
and
$$
b_{11} = b_{22} = b_{33} = b_{42} = \i.
$$
Here $U_c^{\prime} = \d U_c /\d x$.

In equation (\ref{npe}), we confirm that if we 
set $a$, $U_c'$ and $\partial \phi / \partial x$ to zero, we get the parallel 
stability equations of \cite{gill} and \cite{martin}. The boundary 
conditions emerge from requiring that all quantities vary continuously with 
$r$ at the centerline [\cite{gill1}], and obey no-slip at the wall:
\begin{eqnarray}
u =v =w =p =0, \qquad && {\rm at }\  r=0, \ {\rm for }\  n \ne 1, \\
u =p =0, \quad v + \i w = 0, \qquad && {\rm at }\  r=0,\  {\rm for }\  n=1, \\
u = v = w = 0, \qquad && {\rm at }\  r=1. 
\end{eqnarray}
Note that for $n=1$, we have only six boundary conditions for a seventh order 
system. We therefore generate an extra boundary condition by differentiating 
the continuity equation with respect to $r$, and using the fact that 
$u(x,0)=0$, to get 
\begin{equation}
2 {\partial v \over \partial r} + \i n {\partial w \over \partial r}=0 
\qquad  {\rm at }\  r=0,\  {\rm for }\  n=1.
\label{lastofset}
\end{equation}

Equation (\ref{npe}) may be solved as an eigenvalue problem of larger size
[Balachandar \& Govindarajan, preprint, 2005] as described below.
The streamwise derivative in equation (\ref{npe}) couples neighboring axial
locations in the flow-field to one another. Consider two streamwise
locations $1$ and $2$ separated by an incremental distance, i.e., 
$x_2=x_1+\Delta x$. We may write
\begin{equation}
{\partial \phi \over \partial x} = {({{\phi_2}-{\phi_1})} \over \Delta x}+ 
O(\Delta x)^2,
\label{discre}
\end{equation}
Since the dimensional freqency $\beta_d$ remains constant, $\beta_1$ and 
$\beta_2$ are related as follows
\begin{equation}
\kappa \equiv {\beta_2 \over \beta_1} = [1+a \Delta x] {U_{c1} \over U_{c2}}.
\label{relation}
\end{equation}
We can therefore write (\ref{npe}) as 
\begin{equation}
{
\left[ \begin{array}{cc}
   {\cal H}_{1}- {{\cal G}_{1} / \Delta x} & {{\cal G}_{1} / 
\Delta x} \\ {-{\cal G}_{2} / \Delta x} & 
{\cal H}_{2}+ {{\cal G}_{2} / \Delta x}
      \end{array} \right] \\
\left[ \begin{array}{cc}
    \phi_1  \\ \phi_2
      \end{array} \right] \\
=
    \beta_1  
\left[ \begin{array}{cc}
    {\cal B}_{1} & 0 \\ 0 & \kappa {\cal B}_{2}
      \end{array} \right] \\
\left[ \begin{array}{cc}
    \phi_1  \\ \phi_2
      \end{array} \right]. \\
}
\label{array}
\end{equation}
(Higher-order approximations to the streamwise derivative could be considered 
instead of (\ref{discre}) and the resulting 
eigensystem would be of correspondingly large size.) 
The numerical mean flow is interpolated to obtain profiles at neighbouring
$x$-locations, with $\Delta x=0.05$. Profiles obtained from computations using
512 grid points as well as 1024 grid points have been checked to give
eigenvaules correct up to 4 decimal places. Halving or doubling the $\Delta x$
has even less of an effect on the eigenvalue.
Equation (\ref{array})
is solved by a spectral collocation method [\cite{canuto}]. 
The eigenvalue $\beta_1$
is obtained as a complex quantity. The complex streamwise wavenumber is
iterated until $\beta_1$ assumes the desired real value ($=\beta_d R / U_c$)
at a given $x_1$. The axial variation of the wavenumber, $\d\alpha /\d x$, and
the initial guess for $\alpha_1$ are obtained by solving the
homogeneous part of the equation (\ref{npe}). In subsequent iterations, 
$\d\alpha /\d x$ is maintained constant, since the correction due to the
inhomogeneous terms on this quantity is of higher order. It is to be noted
that the apportionment in (\ref{normal_mode}) between the $x$-dependences of 
$\alpha$ and the eigenfunction is arbitrary. 
There are many ways of performing this apportionment, [\cite{Bertolotti}], 
and so long as the rapid (wavelike) change is included in $\alpha_r$, there 
is no difference in the prediction of the growth of any physical quantity.
We have checked that this is the case for the present flow as well.
Choosing $\d\alpha /\d x$ from the homogeneous part of the equation is one
way of including the rapid change into $\alpha(x)$, leaving a relatively 
slow change in $u(x)$.

We consider downstream growth of disturbances followed at a constant value 
of the non-dimensional radius $r$. 
The growth rate $g$ of the nondimensional disturbance kinetic energy, 
$\hat E=1/2(\hat u^2 + \hat v^2 + \hat w^2)$, for example, is given by
\begin{equation}
g = {1 \over \hat E} {\partial \hat E \over \partial x} = -2{\alpha_i} + 
{1 \over E} {\partial E \over \partial x}\Big |_r,
\label{ampf}
\end{equation}
where $E = 1/2 (u u^* + v v^* + w w^*)$, the star denotes a complex 
conjugate. The growth factor for this quantity is thus
\begin{equation}
{E \over E_{cr}} = \exp\left[\int_{x_{cr}}^x g(x) \d x\right]
\label{amplit}
\end{equation}
where the subscript $cr$ stands for the critical location, at which $g=0$.
We see that a disturbance may amplify at one $r$ and decay at another. 
Secondly, one disturbance quantity could be amplifying while others
decay.

\section{Results and discussion}
The slowest decaying mode in a straight pipe is the swirl mode (of 
azimuthal wave number $n=1$) [\cite*{corcos,drazin}]. In the diverging pipe,
we find that this mode again is always the most unstable, all results 
presented here are for $n=1$. We emphasise that for the $3^\circ$ divergence,
a non-parallel stability
analysis is necessary: with a parallel flow assumption, there is no
instability until a Reynolds number of about $1000$.

\begin{figure}
\centering
\begin{minipage}{1.0\textwidth}
\includegraphics[width=0.45\textwidth]{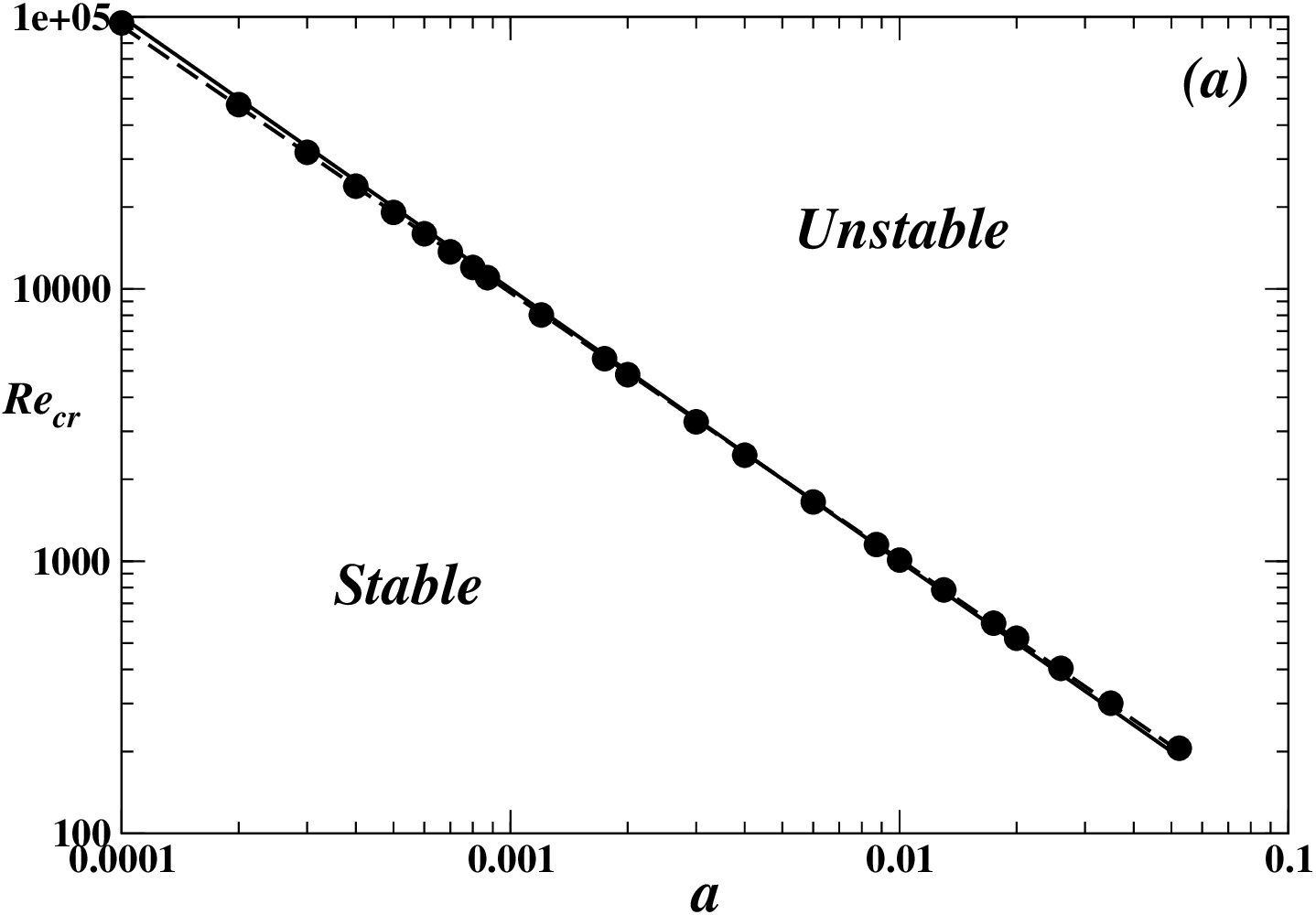}
\hskip 5mm
\includegraphics[width=0.44\textwidth]{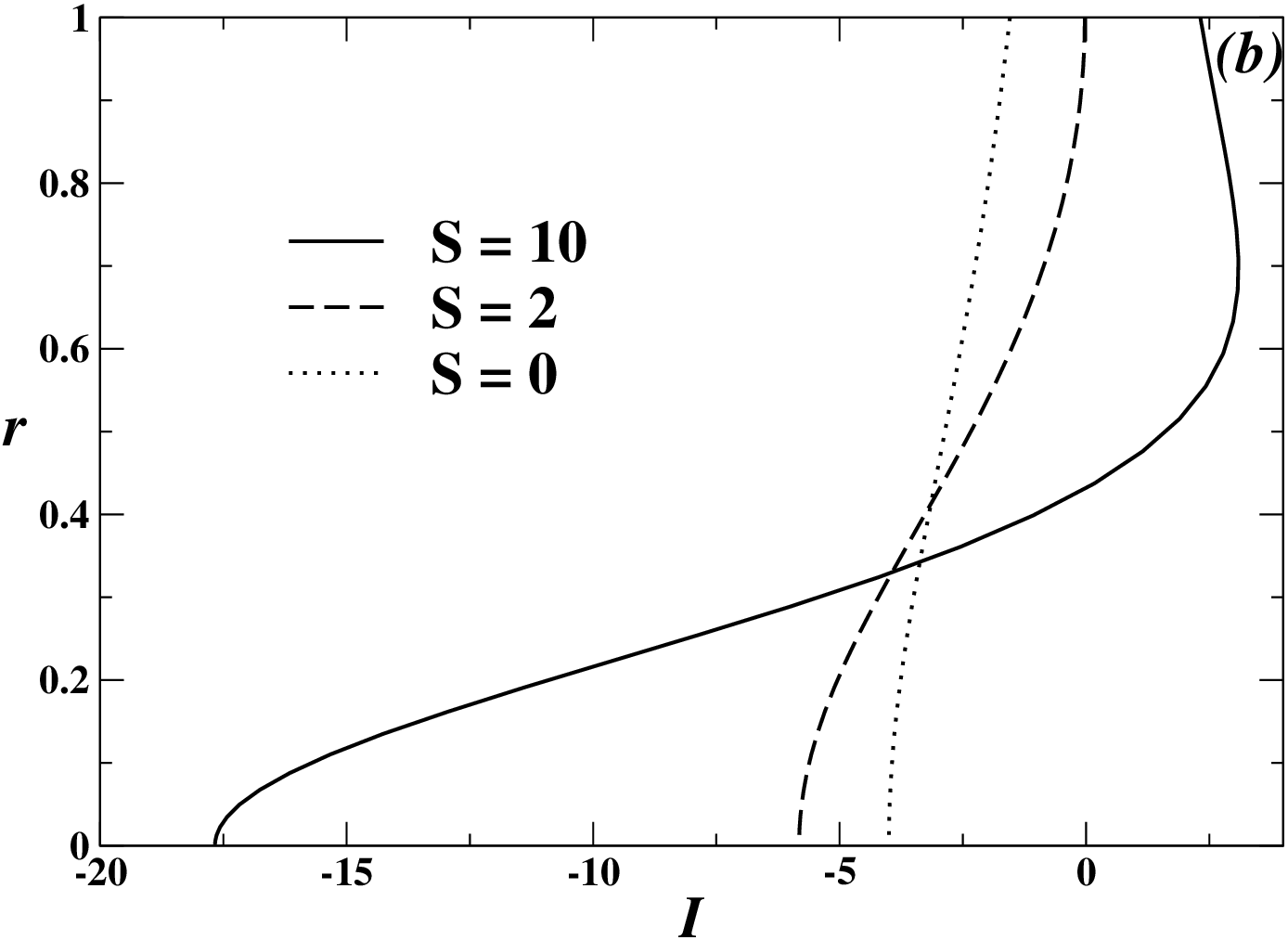}
\caption{(a) Variation of the critical Reynolds number with the tangent $a$ of the
divergence angle, at small angles of divergence. Symbols: stability analysis;
dashed line: best fit; solid line: 
$Re_{cr}=10/a$. (b) The inviscid instability function $I$, in straight 
pipe flow ($S=0$) and for other values of $S$.}
\label{arecr}
\end{minipage}
\end{figure}

We first compare our eigenspectrum for the flow
through a straight pipe with that of \cite{schmid}. Every eigenvalue
matches up to the $10^{th}$ decimal place. Next we study the effect of very
small divergence on the stability behaviour by conducting a parallel flow
stability analysis (neglecting non-parallel terms in the stability equation)
on the AJH profiles. For these parameters, the results have been checked to
be the same as those from a non-parallel analysis.
Figure \ref{arecr}(a) shows the critical Reynolds number for linear instability
as a function of the angle of divergence. At small (but non-zero) divergence 
angle, we find a {\em finite} Reynolds number for linear instability. It can be seen 
that a power-law relationship is obeyed. The best fit of the data gives
$Re_{cr}=11.2 a^{-0.98}$, which is practically indistinguishable from 
$Re_{cr}=10/a$. The critical Reynolds number thus varies as the inverse of the 
divergence angle. The inverse relationship arises because 
an inviscid mechanism is operational at very high Reynolds
numbers, and because the AJH velocity profiles (\ref{similarity})
are described by the product $S$ of $Re$ and $a$. We thus show that there is 
a limiting velocity profile corresponding to $S=10$, and flows where $S$ is
greater than $10$ are linearly unstable. When the angle of divergence is less
than a degree, the AJH profiles are a good approximation of the flow.

To understand this behaviour better, we write down the equation for inviscid
stability (the axisymmetric Rayleigh equation) as the divergence goes to
zero, by neglecting terms
containing $Re^{-1}$ and $a$ in (\ref{npe}), setting $n=1$, and eliminating 
$w$, $p$ and $u$ in turn to get
\begin{equation}
(U-c)\left[v^{\prime\prime} +{3 + \alpha^2 r^2 \over 1+\alpha^2 r^2}
\left({v^\prime \over r}- \alpha^2 v\right) \right]+ \left[U'' - {\alpha^2 r^2-1 \over
r(1+\alpha^2 r^2)}U^\prime\right] v = 0.
\label{quant}
\end{equation}
As $r \to \infty$, the above equation, as it should, reduces to the
two-dimensional Rayleigh equation. In two-dimensional flow, Rayleigh showed
that a necessary condition for instability is that $U''$ goes through a zero
somewhere in the flow [\cite{schmid}]. Several improvements on this 
criterion have been made for
two-dimensional flow, but there is no proof, as far as we know, of a
corresponding necessary condition for pipe flow. We may however follow a
heuristic approach. The quantity $I \equiv U'' - (\alpha^2 r^2-1)/r/(1+
\alpha^2 r^2)U'$ is the axisymmetric analogue in (\ref{quant}) of $U''$ in 
two-dimensional flow. It may therefore be expected that a
change of sign in $I$ within the flow would take the flow
closer to instability. Figure
\ref{arecr}(b) shows the variation of $I$ with the radial coordinate, $I$
undergoes a sign change if $S>2$, consistent with expectation. A value of 
$\alpha=1.26$, corresponding to critical instability, has been used.

\begin{figure}
\centering
\begin{minipage}{1.0\textwidth}
\includegraphics[width=0.48\textwidth]{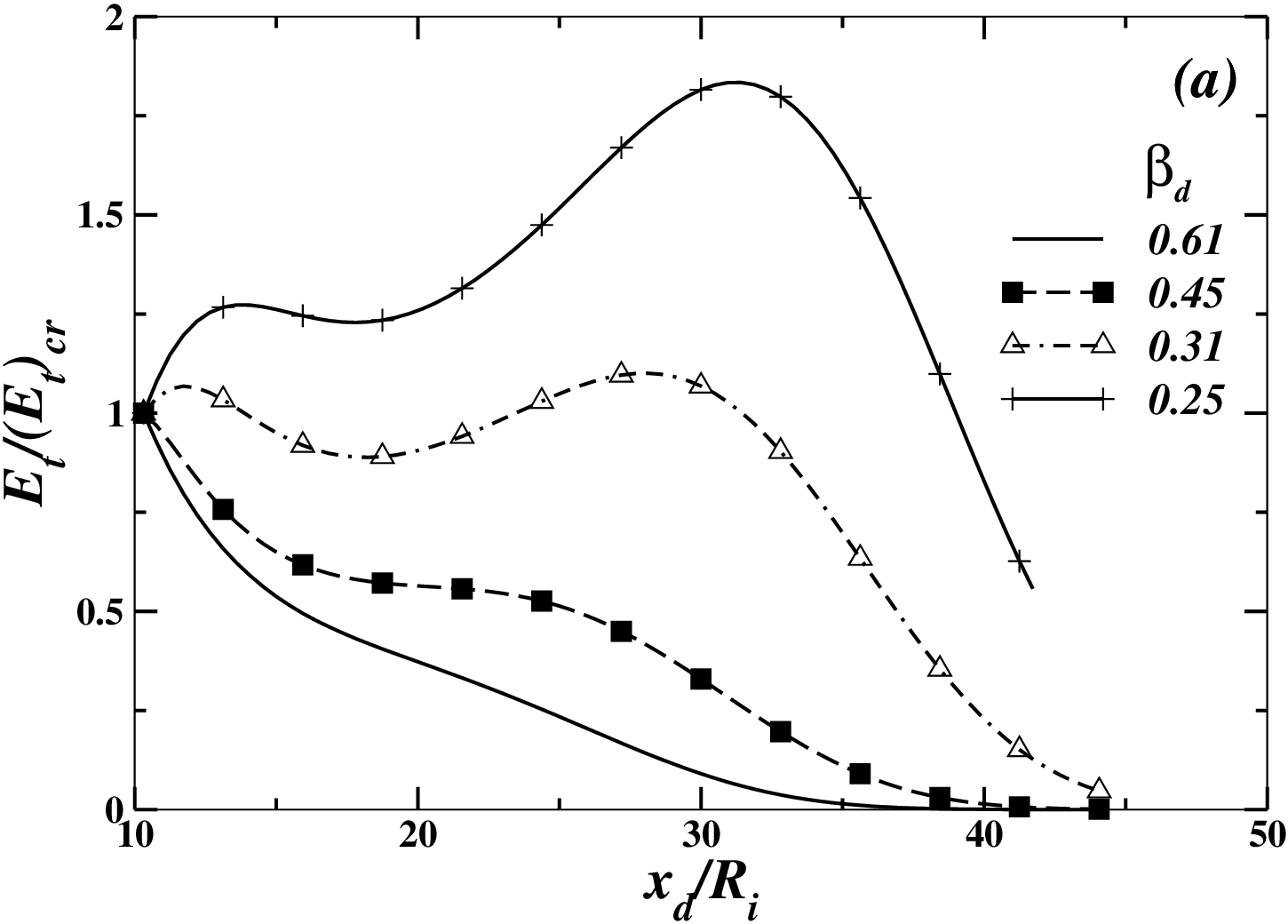}
\hspace{2mm}
\includegraphics[width=0.48\textwidth]{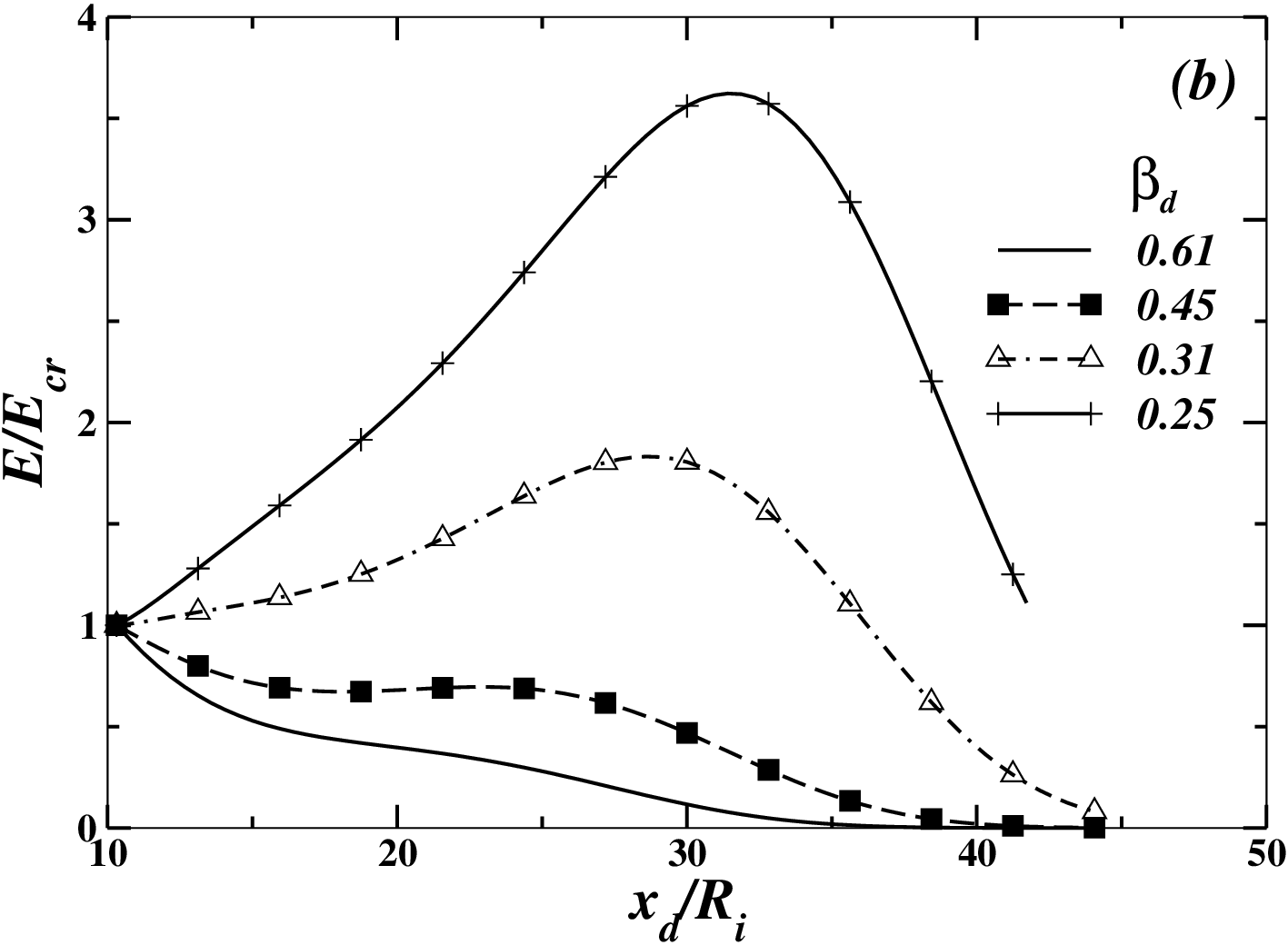}
\includegraphics[width=0.48\textwidth]{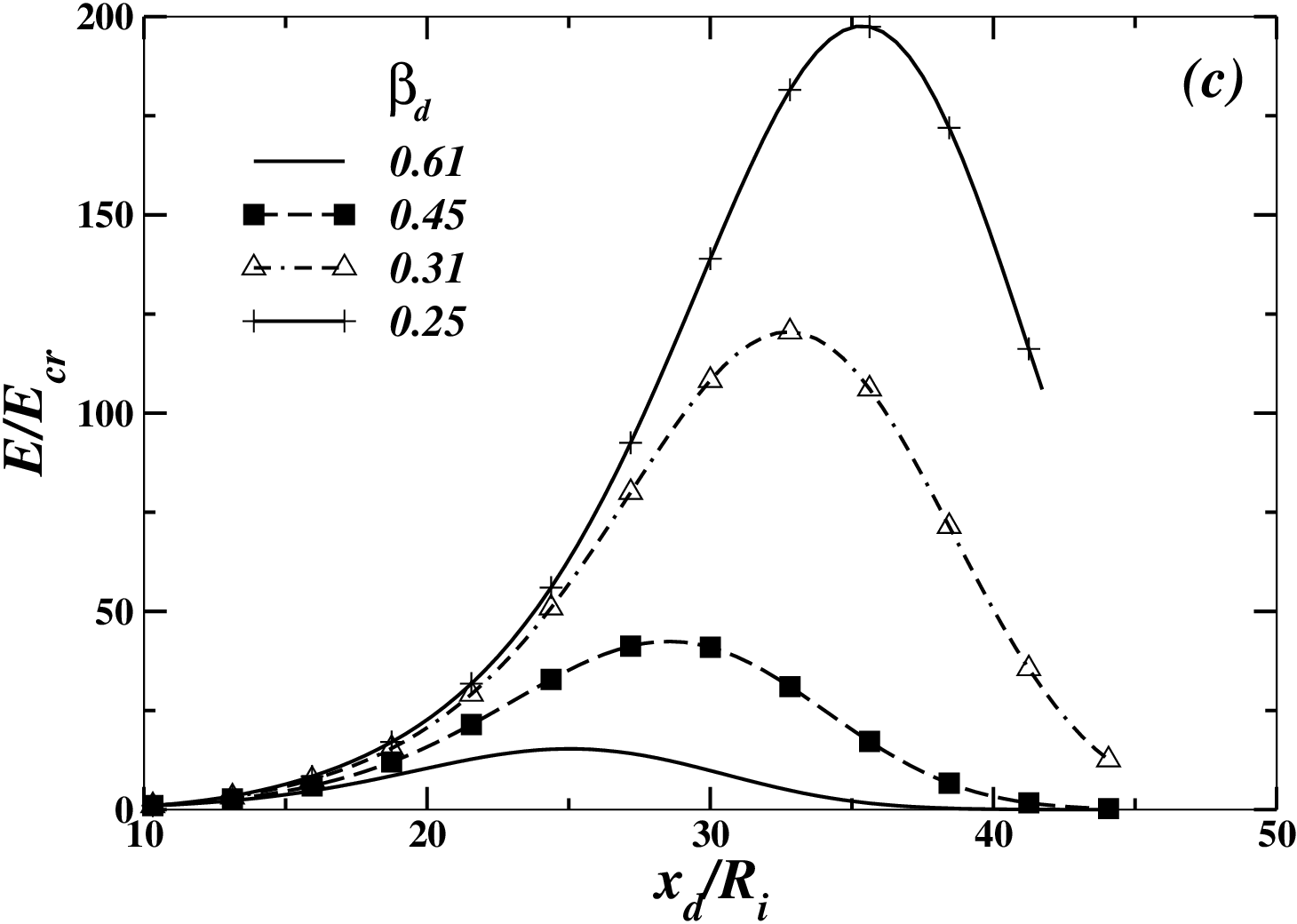}
\hspace{2mm}
\includegraphics[width=0.48\textwidth]{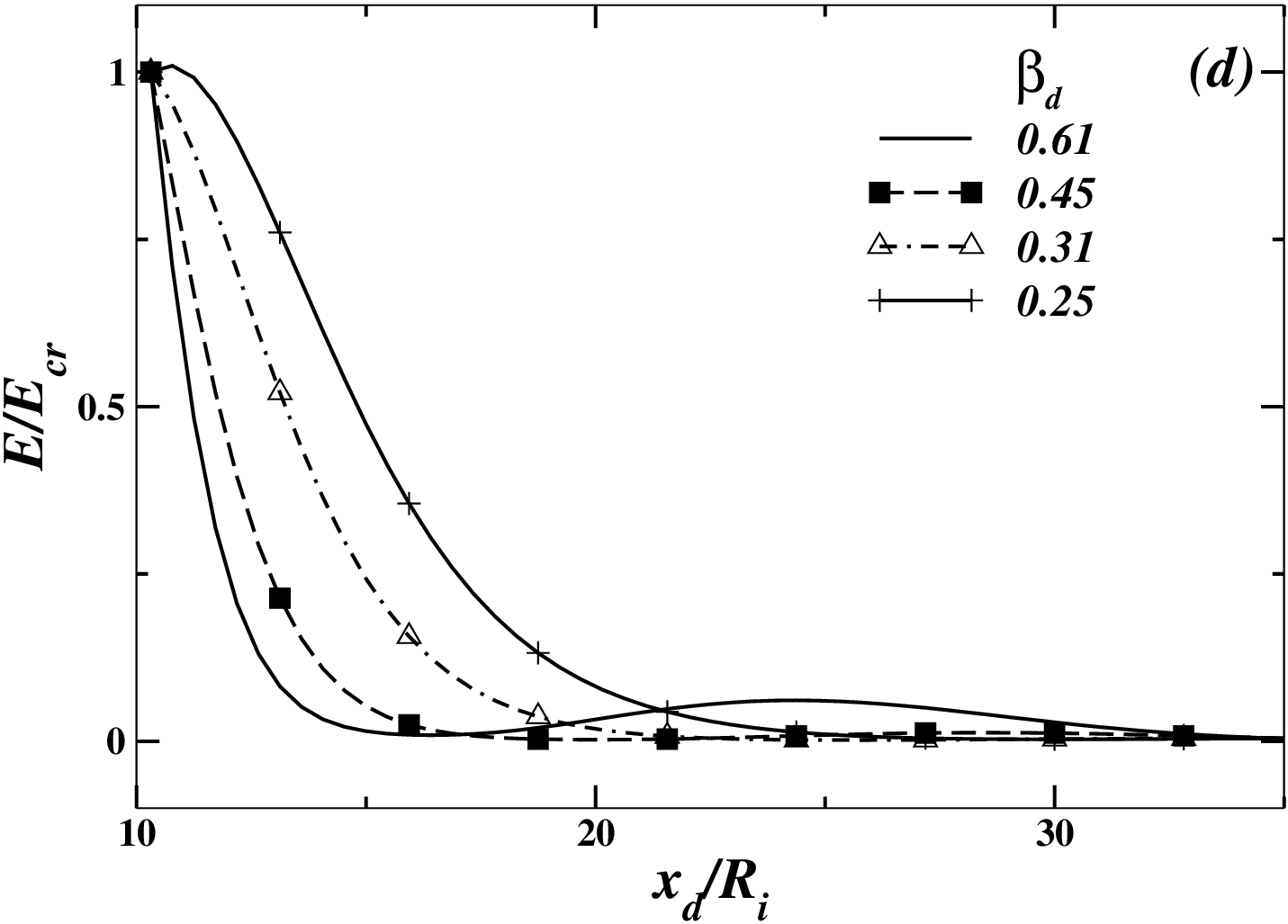}
\caption{Amplification of disturbance kinetic energy for $Re=150$, $n=1$
for typical disturbance frequencies.
(a) Average across the pipe; (b) $r=0.75$; (c) at $r=0.25$;
(d) at $r=0.07$. The axial coordinate here is scaled by the inlet radius.}
\label{growth}
\end{minipage}
\end{figure}

We now examine the behaviour at higher levels of divergence.
For the geometry shown in figure \ref{geometry}, results from
a non-parallel spatial stability analysis [equation (\ref{array})] performed on
the numerically obtained profiles are presented for $Re=150$ 
at the inlet.
The growth rate, as mentioned before, depends on how far the monitoring
location is from the centerline, and what the quantity being monitored is.
An examination of equation (\ref{ampf}) shows that the second term
on the right hand side determines the $r$-dependence, and
comes from the flow quantity under consideration.
\begin{figure}
\centering
\begin{minipage}{1.0\textwidth}
\includegraphics[width=0.45\textwidth]{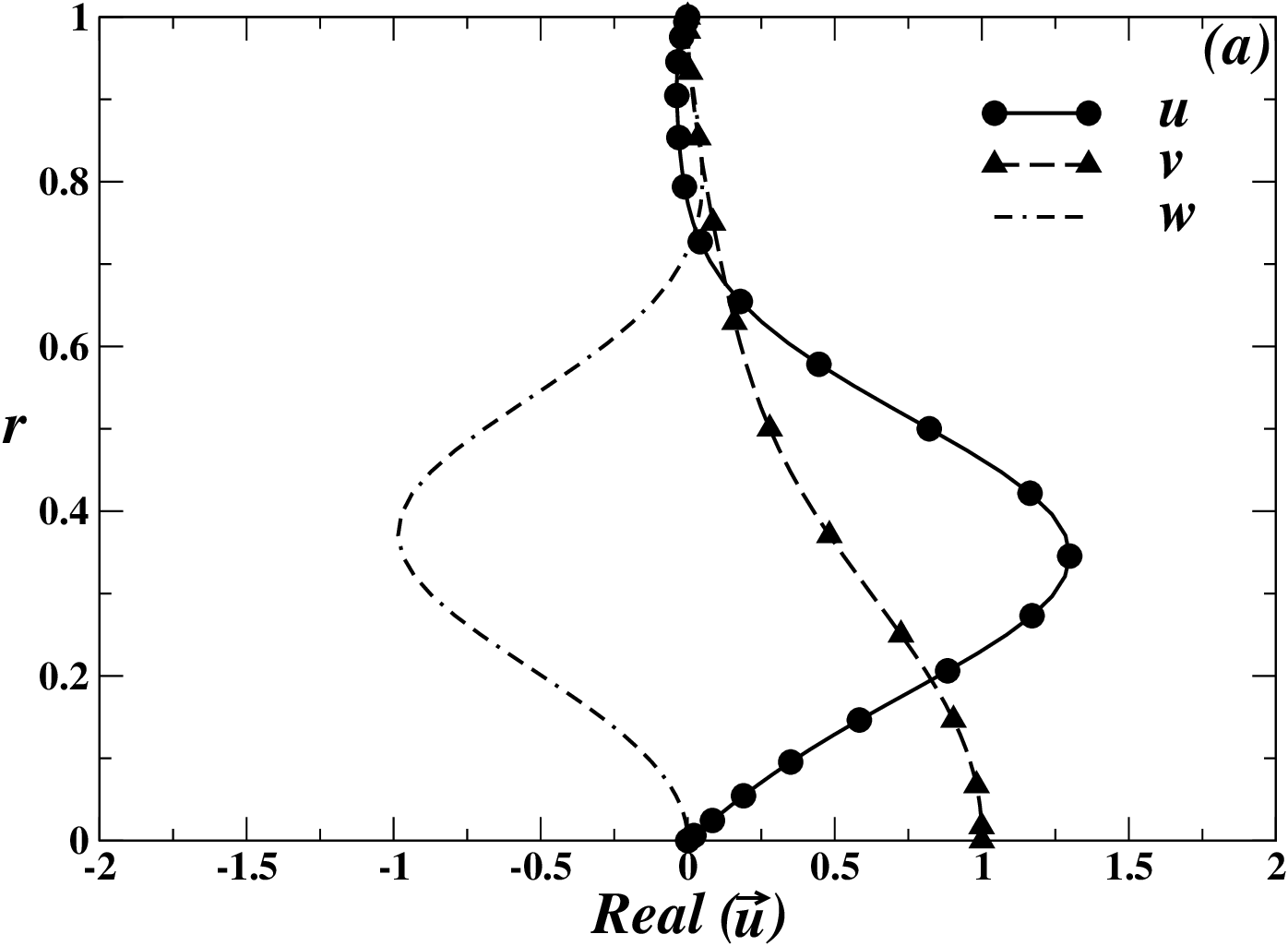}
\hspace{5mm}
\includegraphics[width=0.45\textwidth]{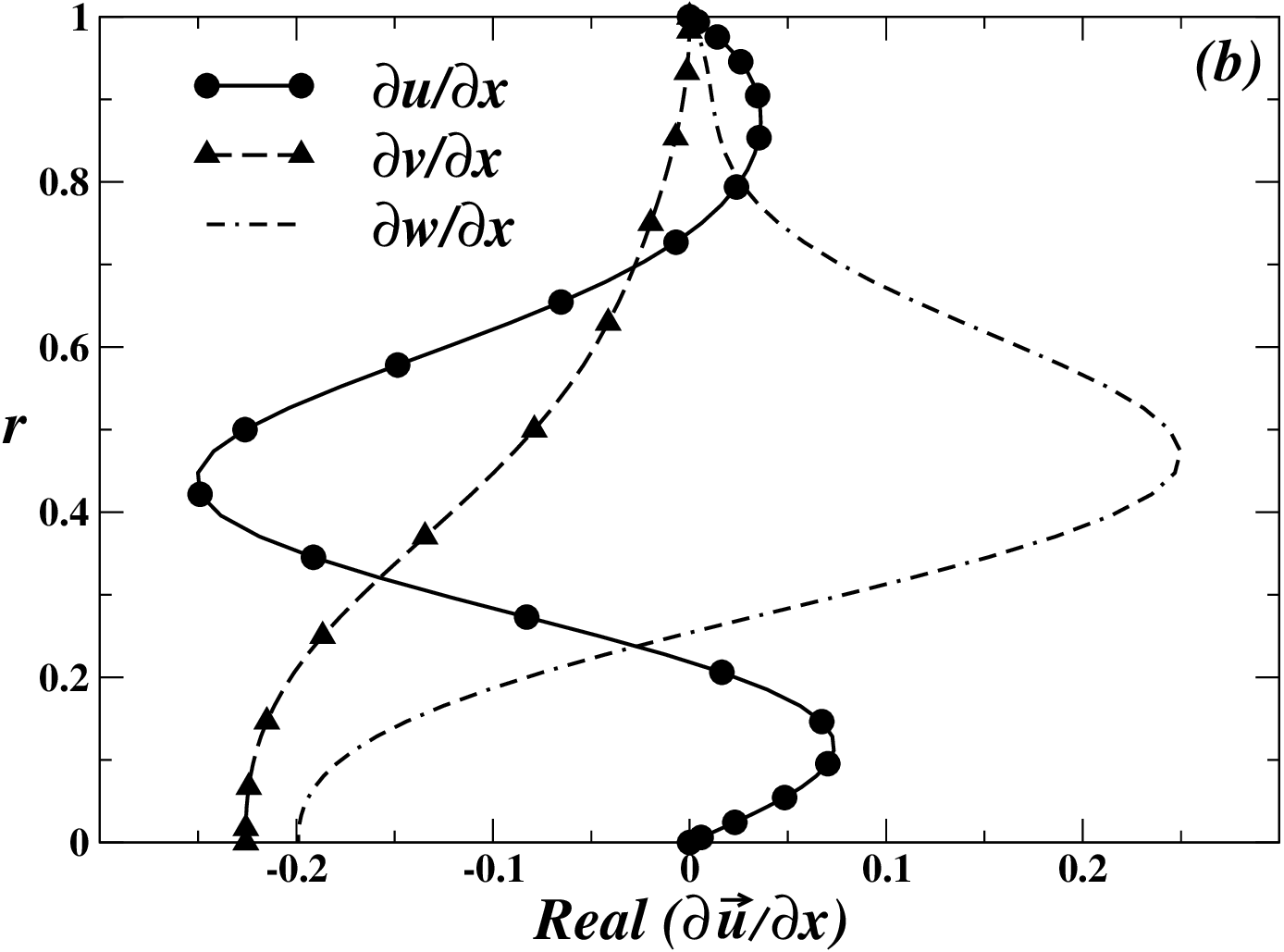}
\caption{(a) Eigenfunctions $\vec u= [u, v, w] $ and (b) ${\partial 
{\vec u} / \partial x}$ for $Re_i = 150$, ${\beta_d}= 0.31$ and $n=1$ 
at $x_d/R_i=28.1$.}
\label{uf}
\end{minipage}
\end{figure}
Typical plots of the eigenfunctions $u$, $v$ and $w$, and their axial
variations, are shown in figures \ref{uf}(a) and (b) respectively.
The amplitude of the disturbance kinetic energy of the swirl ($n=1$) mode
is shown in figure \ref{growth}. It is seen that the fastest growth occurs at
$r=0.25$, while the disturbance decays near the wall and near the centerline.
The Reynolds number is a decreasing function of the axial distance, beyond
$x \sim 50$ we find no disturbance that has a positive growth rate. At higher
Reynolds numbers, there is scope for turbulent flow in the initial portion of
the divergent section, and relaminarisation downstream. These aspects are 
being explored experimentally (Sood {\em et al.}, private communication).
The sensitivity to small levels of divergence may have implications 
for small scale flows.

>From this study, we cannot tell at what Reynolds number transition to
turbulence will be triggered. While an inlet $Re$ of $150$ may be too low, 
it is significant that linear growth has been demonstrated. At $Re=300$ on
the other hand, the disturbance kinetic energy grows by a factor of over 
$60000$ so a linear mechanism may become important in the transition 
process. For a continuously diverging pipe, since the Reynolds number is a 
decreasing function of distance, a 
relaminarisation may occur somewhere downstream, so turbulence could be 
spatially localised.

In summary, the critical Reynolds number for linear instability of flow in
a diverging pipe is finite, and can be surprisingly low, with significant 
disturbance growth rates. The fact that the swirl mode grows exponentially
indicates that a different route to turbulence from that in a straight pipe 
is likely. The highest growth occurs neither close to the wall nor to the 
centreline but in-between. Whether this would give rise to structures 
different from those obtained by \cite{hof04} needs to be investigated.
At divergences as low as $1-2^\circ$, the effect of
flow non-parallelism is already large, so a non-parallel analysis is 
essential. 
As the divergence angle goes to zero, the critical Reynolds number approaches 
infinity as $1/a$.
This instability is generated by an inviscid mechanism.

Grateful thanks are due to Ajay Sood and
Narayanan Menon for useful discussions. We thank the Defence R\&D 
Organisation, Government of India for financial support.

\end{document}